\documentstyle[amstex,aps,epsfig,preprint]{revtex}

\begin{document}

\title{Production of excited electrons at TESLA and CLIC
based $e\gamma$ colliders}
\author{Z. K\i rca$^a$, O. \c{C}ak\i r$^b$ and Z.Z. Ayd\i n$^c$}
\address{$^a$Osmangazi University, Faculty of Arts and Sciences,\\
Department of Physics, 26480, Meselik, Eskisehir, Turkey.\\
$^b$Ankara University, Faculty of Sciences,\\
Department of Physics, 06100, Tandogan, Ankara, Turkey.\\
$^c$Ankara University, Faculty of Engineering,\\ Department of
Engineering Physics, 06100, Tandogan, Ankara, Turkey.}

\maketitle

\begin{abstract}
We analyze the potential of  TESLA and CLIC based electron-photon
colliders to search for excited spin-1/2 electrons. The production
of excited electrons in the resonance channel through the
electron-photon collision and their subsequent decays to leptons
and electroweak gauge bosons are investigated. We study in detail
the three signal channels of excited electrons and the
corresponding backgrounds through
the reactions $e\gamma\rightarrow e\gamma$, $e\gamma\rightarrow eZ$ and $%
e\gamma\rightarrow \nu W$. Excited electrons with masses up to
about $90\%$ of the available collider energy can be probed down
to the coupling $f=f^{\prime}=0.05(0.1)$  at TESLA(CLIC) based
$e\gamma$ colliders.
\end{abstract}

\section{Introduction}
In order to explain the fundamental aspects of the standard model
(SM) such as the number of fermion generations and fermion mass
spectrum compositeness models are expected to be a good candidate.
The replication of three fermionic generations of known quarks and
leptons implies composite structures made up of more fundamental
constituents. The existence of such quark and lepton substructure
leads one to expect a rich spectrum of new particles with unusual
quantum numbers. A possible signal of excited states of quarks and
leptons as predicted by composite models \cite{1,2} would supply
convincing evidence for a new substructure of matter. All
composite models of fermions have an underlying substructure which
is characterized by a scale $\Lambda$. In such models, the known
light fermions would be the ground state spectrum of the excited
fermions.

In order to have an agreement between the precise measurements of
electron and muon $g-2$ and theoretical predictions for chiral
couplings, the compositeness scale $\Lambda$ is expected to be
less than 10 TeV \cite{3}. The absence of electron and muon
electric dipole moments implies the chiral properties of the
excited leptons. A right- handed excited lepton should couple to
only left- handed components of the corresponding lepton. Excited
leptons may be classified by $SU(2)\times U(1)$ quantum numbers
and they are assumed to be both left- and right-handed weak
isodoublets.

Experimental lower limits for the excited electron mass are given
as $m_\star>200$ GeV in \cite{4}, and $m_\star>306$ GeV in
\cite{5}. The higher limits are derived from indirect effects due
to $e^\star$ exchange in the t-channel and depend on transition
magnetic coupling between $e$ and $e^\star$. Relatively small
limits ($m_{\mu^\star,\tau^\star}>94.2$ GeV) for excited muon
($\mu^\star$) and excited tau ($\tau^\star$) are given by the LEP
L3 experiment \cite{6}.

Excited leptons have been studied at $\gamma\gamma$ and $e\gamma$
colliders \cite{6_1}, $e^{+}e^{-}$ colliders
\cite{6_2},\cite{6_3}, and hadron colliders \cite{6_4}.

In this work, resonant production of excited electron in the
s-channel and its subsequent decay modes $e^\star \rightarrow e
\gamma$, $e^\star \rightarrow \nu W$, $e^\star \rightarrow eZ$ are
considered. In addition, we include the contributions coming from
the excited electron in the t-channel. In order to probe excited
electrons, we examine the potential of TESLA and CLIC based
$e\gamma$ colliders with the main parameters given in Table
\ref{table1}. The production cross section and decays of excited
electrons are calculated using an effective Lagrangian which
depends on a compsiteness scale $\Lambda$ and on free parameters
$f$ and $f^{\prime}$.

\section{Effective Lagrangian}

The Lagrangian describing the transition between ordinary and
excited leptons should respect to chiral symmetry in order to
protect the light leptons from acquiring radiatively a large
anomalous magnetic moment. The excited leptons ($l^{\star}$) can
couple to leptons ($l$) and electroweak gauge bosons through the
$SU(2)\times U(1)$ invariant effective interaction Lagrangian
\cite{2}
\begin{equation}
L=\frac{1}{2\Lambda }\overline{l}^{\star }
 \sigma _{\mu \nu
}\left(gf\frac{ \tau }{2}\cdot W_{\mu \nu }+g^{\prime }f^{\prime
}\frac{Y}{2}B_{\mu \nu }\right) l_{L}+\text{H.c.}
\end{equation}
where the $W_{\mu \nu }$ and $B_{\mu \nu }$ represent the field
strength tensors of SU(2) and U(1) gauge fields. The $\tau$
and
$Y$ are the corresponding gauge group generators; $g$ and
$g^{\prime }$ are gauge coupling constants. The parameters $f$ and
$f^{\prime }$ associated to the gauge groups SU(2) and U(1) depend
on compositeness dynamics and they describe the effective changes
from the SM coupling constants $g$ and $g^{\prime }$. In the
physical basis the Lagrangian (1) can be rewritten in more
explicit form

\begin{equation}
L=\frac{g_{e}}{2\Lambda }\left[ (f-f^{\prime })N_{\mu \nu }\sum_{l=e,\nu }%
\overline{l}^{\star }\sigma ^{\mu \nu }l_{L}+f\sum_{l,l^{\prime
}=e,\nu }\Theta _{\mu \nu }^{{\overline l^{\star
}},l}\overline{l}^{\star }\sigma ^{\mu \nu }l_{L}^{\prime }\right]
+\text{H.c.}
\end{equation}
where the first term in the paranthesis is a purely diagonal
$U(1)$ term and vanishes for the coupling $f=f^{\prime }.$ It
contains only triple vertices with
\begin{equation}
N_{\mu \nu }=\partial _{\mu }A_{\nu }-\tan\theta_{W}\partial _{\mu
}Z_{\nu }.
\end{equation}
The second term in (2) is a non-abelian part which involves triple
as well as quartic terms with
\begin{eqnarray}
\Theta _{\mu \nu }^{{\overline\nu ^{\star }},\nu }
&=&\frac{1}{\cos \theta _{W}\sin \theta _{W}}\partial _{\mu
}Z_{\nu }-i\frac{g_{e}}{\sin ^{2}\theta _{W}} W_{\mu }^{+}W_{\nu
}^{-} \\ \Theta _{\mu \nu }^{{\overline e^{\star }},e} &=&-\left(
2\partial _{\mu }A_{\nu }+\frac{ \cos ^{2}\theta _{W}-\sin
^{2}\theta _{W}}{\cos \theta _{W}\sin \theta _{W}}
\partial _{\mu }Z_{\nu }-i\frac{g_{e}}{\sin ^{2}\theta _{W}}W_{\mu
}^{+}W_{\nu }^{-}\right) \\ \Theta _{\mu \nu }^{{\overline\nu
^{\star }},e} &=&\frac{\sqrt{2}}{\sin \theta _{W}} \left( \partial
_{\mu }W_{\nu }^{+}-ig_{e}W_{\mu }^{+}(A_{\nu
}+\cot\theta_{W}Z_{\nu })\right)
\\ \Theta _{\mu \nu }^{{\overline e^{\star }},\nu } &=&\frac{\sqrt{2}}{\sin
\theta _{W}} \left( \partial _{\mu }W_{\nu }^{-}+ig_{e}W_{\mu
}^{-}(A_{\nu }+\cot\theta_{W}Z_{\nu })\right).
\end{eqnarray}

From Eq. (2), the vertex factor for excited lepton ($l^{\star }$)
interacting with lepton ($l$) and gauge bosons ($V=\gamma,Z,W$)
can be obtained as follows
\begin{equation}
\Gamma _{\mu }^{V{\overline l^{\star }}l}=\frac{g_{e}}{2\Lambda
}q^{\nu }\sigma _{\mu \nu }(1-\gamma _{5})f_{V}
\end{equation}
where $q$ is the gauge boson momentum. The couplings $f_{V}$ are
defined by
\begin{eqnarray}
f_{\gamma } &=&Q_{f}f^{\prime }+I_{3L}(f-f^{\prime }) \\
f_{W} &=&\frac{f}{\sqrt{2}\sin \theta _{W}}\\
f_{Z} &=&\frac{-Q_{f}\sin ^{2}\theta _{W}f^{\prime }+I_{3L}(\cos ^{2}\theta
_{W}f+\sin ^{2}\theta _{W}f^{\prime })}{\cos \theta _{W}\sin \theta _{W}}
\end{eqnarray}
where $Q_{f}$ and $I_{3L}$ are the charge of excited lepton and
the weak isospin, respectively; and $\theta _{W}$ is the weak
mixing angle.

\section{Decay Widths}

Decay widths of excited electrons in the individual channels
$e^{\star }\rightarrow \protect e\gamma$, $e^{\star }\rightarrow
eZ$, and $e^{\star }\rightarrow \nu W$ are given by
\begin{equation}
\Gamma (e^{\star }\rightarrow e \gamma )=\frac{\alpha
}{4}\frac{m_{\star }^{3} }{\Lambda ^{2}}f_{\gamma }^{2}
\end{equation}

\begin{equation}
\Gamma (e^{\star }\rightarrow eZ)=\frac{\alpha }{4}\frac{m_{\star
}^{3}}{ \Lambda ^{2}}f_{Z}^{2}(1-\frac{m_{Z}^{2}}{m_{\star
}^{2}})^{2}(1+\frac{ m_{Z}^{2}}{2m_{\star }^{2}})
\end{equation}

\begin{equation}
\Gamma (e^{\star }\rightarrow \nu W )=\frac{\alpha
}{4}\frac{m_{\star }^{3}}{ \Lambda
^{2}}f_{W}^{2}(1-\frac{m_{W}^{2}}{m_{\star }^{2}})^{2}(1+\frac{
m_{W}^{2}}{2m_{\star }^{2}}).
\end{equation}
For $\ m_{\star }\gg m_{Z,W}$, total decay width of excited
electron is given by
\begin{equation}
\Gamma _{tot}\simeq\frac{\alpha }{4}\frac{m_{\star }^{3}}{\Lambda
^{2}}\bigskip \left[ f_{\gamma }^{2}+f_{W}^{2}+f_{Z}^{2}\right]
\end{equation}
The branching ratios for the excited electron decay channels are
described as follows
\begin{equation}
BR=\frac{\Gamma (e^{\star }\rightarrow lV)}{
 \sum_V\Gamma (e^{\star }\rightarrow lV)}.
\end{equation}
where $l$ is electron or neutrino. We choose the parameters either
$f=f^\prime$ or $f=-f^\prime$ in our calculations in order to
reduce the number of free parameters. For the case $f=f^\prime$
($f=-f^\prime$) the coupling of the photon to excited neutrinos
(electrons) vanishes. We display the decay widths and branching
ratios for excited electrons in Fig. \ref{fig1}. The decay widths
of excited electrons, for the accessible mass range, could be
comparable with the detector resolution at TESLA and CLIC based $e
\gamma $ colliders. As can be seen from Fig. \ref{fig1} an excited
electron decays into a W boson and a neutrino dominantly, and  the
branching ratios are insensitive to higher excited electron mass
when compared to $m_W$ or $m_Z$. We obtain the limiting values for
the branching ratios at large $m_\star$ as 0.28, 0.60 and 0.11 for
the coupling $f=f^{\prime}=1$ at photon, W and Z channel,
respectively. In the case $f=-f^{\prime}=-1$, the branching ratio
for W channel does not change while it increases to the value
$0.39$ for Z channel.

\section{Cross Sections}

Excited electrons can be produced directly via the subprocess
$e\gamma \rightarrow e^{\star}\rightarrow lV$ ($V=\gamma,Z,W$) and
indirectly via t-channel exchange diagram. The Feynman diagrams
for $e\gamma \rightarrow e\gamma(eZ)$ and $\gamma e\rightarrow \nu
W $ processes in electron-photon collisions are shown in Fig.
\ref{fig2} and \ref{fig3}.

For an immediate estimation, the cross section for the signal can be well
approximated with the Breit Wigner formula
\begin{eqnarray}
\hat{\sigma}_{BW} &=&\frac{8\pi ^{2}\Gamma _{i}\cdot \Gamma _{f}}{m_{\star
}s\Gamma _{tot}}f_{\gamma }(x)
\end{eqnarray}
for the narrow decay widths where $\Gamma_{i}$ and $\Gamma_{f}$
are the initial and final state decay widths, respectively. Here,
$x=\hat s/s$ where $\sqrt{\hat s}$ being the center of mass energy
of the subprocess. In order to obtain the total cross sections for
the signal and background, without the narrow width approximation,
we use the following formula
\begin{eqnarray} \sigma =\int_{x_{\min
}}^{0.83}dxf_{\gamma }(x)\widehat{\sigma }(\widehat{s}).
\end{eqnarray}
where $\widehat{\sigma }(\widehat{s} )$ is obtained from the well-
known matrix element calculation from the Feynman diagrams given
in Figs. \ref{fig2} and \ref{fig3}. Here, $x_{\min
}={m_{\star}^2}/{s}$. The high energy photon spectrum $f_{\gamma
}(x)$ obtained from the Compton backscattering is given by

\begin{eqnarray}
f_{\gamma}(x)=\left\{
\begin{array}{cc}
\frac{1}{N}\left[ 1-x+\frac{1}{1-x}\left[
1-\frac{4x}{x_{0}}\left( 1-\frac{x}{x_{0}\left( 1-x\right) }\right)
\right] \right] & ,0<x<x_{\max } \\
0 & ,x>x_{\max }
\end{array}\right\}
\end{eqnarray}
where $x_{0}=4.82$, $x_{max}={x_{0}}/{(1+x_{0})}$ and N=1.84
\cite{7}. The production cross section of excited electron in
three modes, taking $f=f^\prime=1$ and $\Lambda$=$m_\star$, are
given in Fig. \ref{fig4} and \ref{fig5} for TESLA and CLIC based
$e\gamma$ colliders at the center of mass energies of
$\sqrt{s}=911$ GeV and $\sqrt{s}=2733$ GeV, respectively. From
Fig. \ref{fig4} and Fig. \ref{fig5}, we get the following
information; the $\nu W$ channel gives higher cross section than
the others however there is an ambiguity with the neutrino in this
channel. Therefore, the photon channel gives a more promising
result because of its simple kinematics. Excited electrons can be
produced copiously at TESLA and CLIC based $e\gamma $ colliders.
The cross sections and the numbers of signal events are shown in
Table \ref{table2} and \ref{table3}. For the signal and background
processes we apply a cut $p^{e,\gamma}_T > 10$ GeV for
experimental identification of final state particles. The
backgrounds to the W and Z decay channels in the hadronic final
states are fairly large. The backgrounds to the photonic final
states are relatively small in comparasion with the W channels.

In order to get one particle inclusive cross sections for the
production of a particle of transverse momentum $p_{T}$ and
rapidity $y$, we use the following standard procedure. The
differential cross section for the process {$e\gamma\rightarrow
eZ$} with respect to the transverse momentum $p_T$ of outgoing
electron is given by

\begin{eqnarray}
\frac{d\sigma }{dp_{T}}=2p_{T}\int_{y^{-}}^{y^{+}}dy\ f_{\gamma
/e}(x)\frac{ xs}{\mid s-2p_{T}E_{a}e^{-y}\mid
}\frac{d\widehat{\sigma }}{d\widehat{t}}
\end{eqnarray}
with
\begin{eqnarray}
x=\frac{2p_{T}E_{b}e^{y}+m_{Z}^{2}}{s-2p_{T}E_{a}e^{-y}} \nonumber
\end{eqnarray}
and
\begin{eqnarray}
y^{\pm }=\log \left[ \frac{0.83s-m_{Z}^{2}}{4p_{T}E_{b}}\pm
\sqrt{\left( \frac{0.83s-m_{Z}^{2}}{4p_{T}E_{b}}\right)
^{2}-\frac{0.83E_{a}}{E_{b}}} \right] \nonumber
\end{eqnarray}
where $E_{a}$ and $E_{b}$ are the incoming particle energies.
$\hat{s}$ and $\hat{t}$ are the Lorentz invariant Mandelstam
variables.

The $p_{T}$ distribution of $W$ boson in $e\gamma $ collision at
rapidity $y$ for the process {$e\gamma\rightarrow \nu W$} is given
by

\begin{eqnarray}
\frac{d\sigma }{dp_{T}}=2p_{T}\int_{y^{-}}^{y^{+}}dy\
f_{\gamma}(x)\frac{ xs}{\mid s-2m_{T}E_{a}e^{-y}\mid
}\frac{d\widehat{\sigma }}{d\widehat{t}}
\end{eqnarray} with
\begin{eqnarray}
x=\frac{2m_{T}E_{b}e^{y}-m_{W}^{2}}{s-2m_{T}E_{a}e^{-y}} \nonumber
\end{eqnarray}
and

\begin{eqnarray}
y^{\pm }=\log \left[ \frac{0.83s+m_{W}^{2}}{4m_{T}E_{b}}\pm
\sqrt{\left( \frac{0.83s+m_{W}^{2}}{4m_{T}E_{b}}\right)
^{2}-\frac{0.83E_{a}}{E_{b}}} \right] \nonumber
\end{eqnarray} where
$m_{T}=\sqrt{m_{W}^{2}+p_{T}^{2}}$ is the definition for the
transverse mass of $W$ boson.

The production of a photon in $e\gamma $ collision at rapidity $y$
and transverse momentum $p_{T}$ for the process
{$e\gamma\rightarrow e\gamma $ can be found by replacing $m_W=0$
in (21).

In the Eqs. (20) and (21), we have calculated the differential
cross sections $d\hat{\sigma}/d\hat{t}$ for the signal and
background processes taking into account the interferences between
the SM and excited electrons contributions.

For both signal and background, the behavior of $p_{T}$ spectrum
of final state photon, $W$ boson and electron in three modes are
shown in Figures \ref{fig6}-\ref{fig8} for various values of
parameters $f=f^\prime$ at TESLA based $e\gamma$ collider with
$\sqrt{s}=911$ GeV. At CLIC based $e\gamma$ collider with
$\sqrt{s}=2733$ GeV, we can easily scale the cross sections
according to Figs. \ref{fig4} and \ref{fig5}. For signal process
{$e\gamma \rightarrow e^{\ast }\rightarrow e\gamma $}, transverse
momentum $p_{T}$ distribution of photon or electron is peaked
around the half of the mass value of excited electron. For the
process {$e\gamma \rightarrow e^{\ast }\rightarrow \nu W$}(eZ),
$p_{T}$ distribution of W boson (electron) shows a peak around
${m_{*}/2}-{m^{2}_{V}/2m_{*}}$. Here $m_V$ denotes W boson (or Z
boson) mass. This signal peak moves to a greater (smaller) $p_T$
value when the excited electron mass increased (decreased). For
the parameters $f=f^\prime\neq1$ one can conclude that the $p_{T}$
distribution changes simply with $f^2$. The backgrounds decrease
smoothly when the transverse momentum of final state particles
increase.

In order to estimate the number of events for the signal and
background in a chosen $p_T$ window,
we integrate the transverse momentum distribution
around the half of each excited electron mass point
($m_\star/2-m_V^2/2m_\star$) in the
interval of transverse momentum resolution $\Delta p_T$. Here $\Delta p_T$ is
approximated as $\approx 10$ GeV  for
$m_\star=500$ GeV and $\approx 20$ GeV for $m_\star=1500$ GeV for a
generic detector. In order to
calculate signal significance, we use
the integrated luminosity for TESLA and CLIC based $e \gamma$
colliders with $L=9.4\times 10^4 $ pb$^{-1}$ and $L=9\times 10^4 $
pb$^{-1}$, respectively \cite{8}.

In order to quantify the potential of TESLA and CLIC based $e
\gamma$ colliders to search for excited electron, we define the
statistical significance (SS)
\begin{center}
\begin{eqnarray}
SS&=&\frac{\left| \sigma _{total}-\sigma _{back}\right|}{\sqrt{%
\sigma _{back}}}\sqrt{L}  .
\end{eqnarray}
\end{center}
where $\sigma_{total}$ and $\sigma_{back}$ is the total cross
sections for signal+background and background inside chosen $p_T$
window, respectively. We calculate the value of SS for different
couplings assuming $f=f^\prime$ and requiring the condition $SS>5$
for the signal observability. We find from the Tables
\ref{table4}-\ref{table7} that the excited electrons can be
observed down to coupling $f\simeq 0.05$ at TESLA and $f\simeq
0.1$ at CLIC based $e\gamma$ colliders.

In the final states containing W or Z boson we should consider the
subsequent decay modes of the weak bosons. For leptonic decay the
branching ratios are $BR(W \rightarrow l\nu)\cong 10.56 \% $ and
$BR(Z \rightarrow l^{+}l^{-})\cong 3.37\% $. However hadronic
decay modes have larger branchings as $BR(W \rightarrow
Hadrons)\cong 68.5 \% $ and $BR(Z \rightarrow Hadrons)\cong 69.89
\%$ \cite{9}. We should multiply the cross sections for $2
\rightarrow 2$ processes by the branching ratios for subsequent
decays of W or Z boson in the weak decay channels of excited
electron. Here we do not consider the invisible decay modes. The
direct production of excited electrons at $e\gamma$ collider will
give the signal in the final state a) electron+photon or b)
lepton+$p^{miss}_T$ or c) 2jet+$p^{miss}_T$ or d) electron+2lepton
or e)electron+$p^{miss}_T$ or f) electron+ 2jet. For the
observation of more clear signal we may choose the leptonic
channels.

\section{Results and Discussions}

Excited electrons can be produced directly with a large cross
section (even in the three decay modes) at high energy TESLA and
CLIC based $e\gamma$ colliders. For an observation, we require the
condition $SS>5$ per year at $e\gamma$ colliders with the
integrated luminosity of $O(\sim 10^5 pb^{-1})$. With the
couplings $f=f^{\prime}=1$ we can reach the excited electron mass
up to the kinematical limits of collider energies. For smaller
values of the parameters $f$ and $f^\prime$, the cross sections
are lowered simply by $f^2$. If excited electrons with a lower
parameters exist, we will need high resolution detectors.

In this study, we have taken into account that the excited
electrons interact with the Standard Model particles through the
effective Lagrangian (1). This may be a conservative assumption
because it is possible for excited fermions couple to ordinary
quarks and leptons via contact interactions originating from the
strong constituents dynamics. In this case, the decay widhts can
be enhanced \cite{10}.

In conclusion, we have presented the results of excited electron
production with subsequent decays into three decay channels. The
excited electrons can be produced at TESLA and CLIC based
$e\gamma$ colliders up to kinematical limit at each channel due to
the smooth photon energy spectrum. We find that excited electron
with mass $500(750)$ GeV can be probed down to the coupling
$f=f^{\prime}\simeq 0.05$ at TESLA based $e\gamma$ colliders. At a
CLIC based $e\gamma$ collider ($\sqrt{s}=2733$ GeV), the excited
electron can be probed down to the couplings $f=f^{\prime}\simeq
0.1$.

\newpage

\begin{table}[tbp]
\caption{Main parameters of TESLA and CLIC $e^{+}e^{-}$ colliders and
$e\gamma$ colliders based on them.}
\label{table1}
\begin{tabular}{llll}
\hline &  $\sqrt{s}_{ee}$ (GeV) & $\sqrt{s}_{e\gamma}$ (GeV) &
$L_{e\gamma}(10 ^{34} cm^{-2}sn^{-1})$
\\ \hline
TESLA & 1000 & 911 & 0.94 \\ \hline CLIC& 1000 & 911  & 0.35
\\
\hline CLIC& 3000 & 2733& 0.90
\\
\hline
\end{tabular}
\end{table}

\begin{table}[tbp]
\caption{Cross sections and number of events for signal at
$\sqrt{s_{e\gamma}}=911$ GeV. Here the numbers $N_1$ and $N_2$ are
for TESLA and CLIC based $e\gamma$ colliders, respectively.}
\label{table2}
\begin{tabular}{llllllllll}
\hline & \multicolumn{3}{c}{$e\gamma\rightarrow e\gamma $} &
\multicolumn{3}{c}{$ e\gamma\rightarrow \nu W$} &
\multicolumn{3}{c}{$e\gamma\rightarrow eZ$}
\\ \hline
$m_{\ast }$(GeV) & $\sigma$ (pb)& $N_1(10^2)$ & $ N_2(10^2)$ &
$\sigma$(pb)& $N_1(10^2)$& $N_2(10^2$ )&$\sigma$ ( pb) &
$N_1(10^2)$&$N_2(10^2)$
\\ \hline 200 &27.80
& 26132 &9730 & 38.82 & 36491 & 13587&10.57&9936&3699\\ \hline 300
&14.22 & 13367 &4977& 26.11 &24543 & 9139& 5.32 & 5000&1862\\
\hline 400 &11.50 & 10810 &4025& 23.40 & 21996&8190 & 4.46 &
4192&1561
\\ \hline 500 &9.87 & 9278&3455& 20.85&19599&7298& 3.92 & 3685&1372\\
\hline 600 &9.22& 8667&3227 & 19.80&18612&6930& 3.71 & 3487&1299
\\ \hline 700 &9.26& 8704  &3241 & 20.06 & 18856&7021&3.76 &
3535& 1316
\\ \hline 800 &11.07 & 10406 & 3875&24.13 &22682&8446& 4.52
&4249&1582
\\ \hline 900 & 20.74 & 19496 &7259& 45.34&42620&15869 &8.49 &
7981&2972
 \\ \hline
\end{tabular}
\end{table}

{\scriptsize
\begin{table}[tbp]
\caption{Cross sections and number of events for signal at CLIC
based $e\gamma$ collider with $\sqrt{s}=2733$ GeV and integrated
luminosity $L=9\times 10^4 $pb$^{-1}$.} \label{table3}
\begin{tabular}{lllllll}
\hline & \multicolumn{2}{c}{$e\gamma\rightarrow e\gamma $} &
\multicolumn{2}{c}{$ e\gamma\rightarrow \nu W $} &
\multicolumn{2}{c}{$e\gamma\rightarrow eZ$}
\\ \hline
$m_{\ast }($GeV$)$ & $\sigma
($pb$)$&N($10^2$)&$\sigma($pb$)$&N($10^2$)& $\sigma ($ pb$)$ &
N($10^2$)\\ \hline 200 & 197.10 & 177400  &223.50 & 201200 & 81.74
& 73570\\ \hline 400 &12.53 &11280 & 15.52  & 13970 & 5.19 &
4671\\ \hline 600 &3.07 & 2763 & 4.64& 4176 &1.27 & 1143  \\
\hline  800 &1.73 & 1557&3.21 &2889  & 0.71 & 639
\\ \hline
1000 &1.41 & 1269 &2.89  & 2601 & 0.58 & 522\\ \hline  1200 &1.23
& 1107 & 2.63 & 2367 & 0.51 & 459  \\ \hline 1400 &1.11 & 999 &
2.41 & 2169  & 0.46 & 414 \\ \hline  1600 &1.04 & 936 & 2.28 &
2052& 0.43 & 387 \\ \hline  1800 &1.01  & 907 &2.22 & 1998 & 0.41
& 369
\\ \hline  2000
&1.00 & 903 & 2.21 &1989  & 0.41 & 369 \\ \hline  2200 &1.05 &
945& 2.31 & 2079 & 0.43 & 387 \\ \hline  2400 &1.22  & 1098 &2.70
& 2430  & 0.50 & 450
\\ \hline  2600
&1.80 & 1620& 3.98 & 3582 & 0.75 & 675 \\ \hline
\end{tabular}
\end{table}}

\begin{table}[tbp]
\caption{The total cross sections of signal and backgrounds inside
the bin chosen and statistical significance (SS) values according
to different $f=f^\prime$ for $m_{\ast }=500$ GeV at
$\sqrt{s}=911$ GeV with $L=9.4\times 10^4$pb$^{-1}$.}
\label{table4}
\begin{tabular}{lllllll}\hline
&  & \multicolumn{3}{c}{$m_{\ast }=500$ GeV} &  \\ \hline
& \multicolumn{1}{c}{$e\gamma\rightarrow e\gamma $} &  & \multicolumn{1}{c}{$
e\gamma\rightarrow \nu W $} &   &
\multicolumn{1}{c}{$e\gamma\rightarrow eZ$} &
\\ \hline
$f=f^{\prime}$ & $\sigma_{tot} ($pb$)$ & $SS$ &
$\sigma_{tot}($pb$)$ & $SS$ &  $\sigma_{tot}
($pb$)$ & $SS$ \\ \hline
1.00& 24.80 & 2100.4 &  78.00 & 1711.3 & 8.50 & 1758.8   \\
\hline
0.50& 11.20 & 508.5 &  50.80 & 422.8 & 3.20 & 424.5  \\
\hline 0.10& 7.03 &19.8 &  41.90& 3.1 &   1.60 & 16.6  \\
\hline 0.05& 6.87 & 0.7   &    &   & 1.52 & 2.9\\ \hline
\end{tabular}
\end{table}

\begin{table}[tbp]
\caption{The total cross sections of signal and background inside
the bin chosen and statistical significance (SS) values according
to different $f=f^\prime$ for $m_{\ast }=750$ GeV at
$\sqrt{s}=911$ GeV with $L=9.4\times 10^4$pb$^{-1}$.}
\label{table5}
\begin{tabular}{lllllll}
\hline &  & \multicolumn{3}{c}{$m_{\ast }=750$ GeV} &  \\ \hline
& \multicolumn{1}{c}{$e\gamma\rightarrow e\gamma$} &  &
\multicolumn{1}{c} {$e\gamma\rightarrow \nu W$} &  &
\multicolumn{1}{c}{$e\gamma\rightarrow eZ$} &
\\ \hline
$f=f^{\prime}$ & $\sigma_{tot} ($pb$)$ & $SS$ &
$\sigma_{tot}($pb$)$ & $SS$ &  $\sigma_{tot}
($pb$)$ & $SS$
\\ \hline
1.00& 24.70 & 1834.8 &  80.10 & 1812.4 & 8.90 & 1847.5  \\
\hline
0.50& 11.30 & 455.3 &  51.40 &452.3 &  3.40 & 673.3 \\
\hline 0.10& 7.04 & 18.4 &  42.13& 12.8 &  1.60& 23.6  \\
\hline 0.05& 6.87 & 0.7 &  41.88 & 1.1 & 1.52 & 3.8   \\ \hline
\end{tabular}
\end{table}

\begin{table}[tbp]
\caption{The total cross sections of signal and background inside
the bin chosen and statistical significance (SS) values according
to different $f=f^\prime$ for $m_{\ast }=750$ GeV at
$\sqrt{s}=2733$ GeV with $L=9\times 10^4$pb$^{-1}$.}
\label{table6}
\begin{tabular}{lllllll}
\hline &  & \multicolumn{3}{c}{$m_{\ast }=750$ GeV} &  \\ \hline
& \multicolumn{1}{c}{$e\gamma\rightarrow e\gamma $} &  & \multicolumn{1}{c}{$
e\gamma\rightarrow \nu W$} &  &
\multicolumn{1}{c}{$e\gamma\rightarrow eZ$} &
\\ \hline $f=f^{\prime}$ & $\sigma_{tot} ($pb$)$ & $SS$ &
$\sigma_{tot}($pb$)$ & $SS$ &  $\sigma_{tot}
($pb$)$ & $SS$
\\ \hline
1.00& 4.30 & 866.8 &  53.40 & 190.2 &  1.51 & 634.9  \\
\hline 0.50& 1.80 & 165.4 &  50.00 & 46.7 & 0.53 & 118.1  \\
\hline 0.10& 1.20 &5.5 &  48.94& 0.6 & 0.32& 3.7  \\
\hline 0.05& 1.18& 0.7 &  & &  &
\\ \hline
\end{tabular}
\end{table}

\begin{table}[tbp]
\caption{The total cross sections of signal and background inside
the bin chosen and statistical significance values according to
different $f=f^\prime$ for $m_{\ast }=1500$ GeV at $\sqrt{s}=2733$
GeV with $L=9\times 10^4$pb$^{-1}$.} \label{table7}
\begin{tabular}{lllllll}
\hline &  & \multicolumn{3}{c}{$m_{\ast }=1500$ GeV} &  \\ \hline
& \multicolumn{1}{c}{$e\gamma\rightarrow e\gamma $} &  & \multicolumn{1}{c}{$
e\gamma\rightarrow \nu W $} &  &
\multicolumn{1}{c}{$e\gamma\rightarrow eZ$} &
\\ \hline
$f=f^{\prime}$ & $\sigma_{tot} ($pb$)$ & $SS$ &
$\sigma_{tot}($pb$)$ & $SS$ &  $\sigma_{tot}
($pb$)$ & $SS$
\\
\hline 1.00& 3.10 & 537.9 &  52.90 & 173.2 & 1.12 & 427.7   \\
\hline 0.50& 1.60 & 130.3 &  49.90 & 43.7 & 0.51 & 102.8 \\
\hline  0.10 & 1.20& 4.3   &  48.93& 0.1 & 0.32 & 2.8  \\ \hline
\end{tabular}
\end{table}

\newpage

\begin{figure}[tbp]
\begin{center}
\epsfig{file=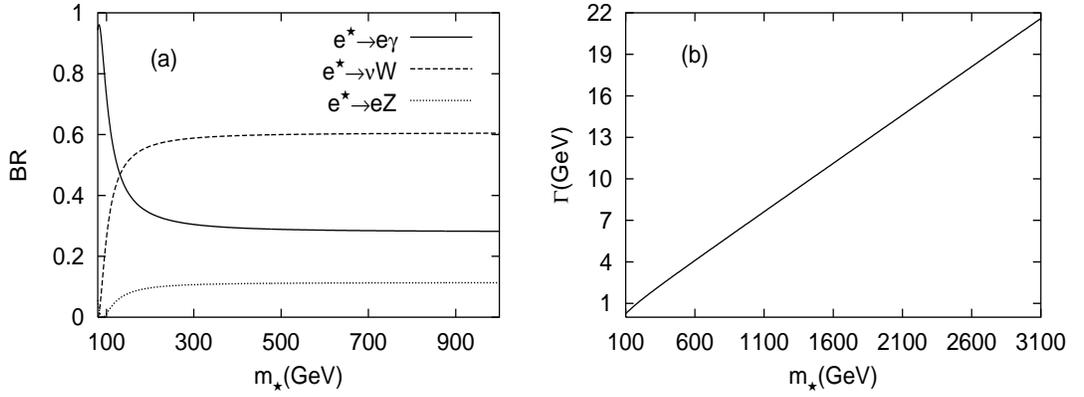,width=12cm,height=9cm}
\end{center}
\caption{The branching ratios (a) and the total decay width (b)
for excited electrons as a function of its mass with
$\Lambda=m_{\star}$ and the coupling $f=f^{\prime}=1$.}
\label{fig1}
\end{figure}

\begin{figure}[tbp]
\begin{center}
\epsfig{file=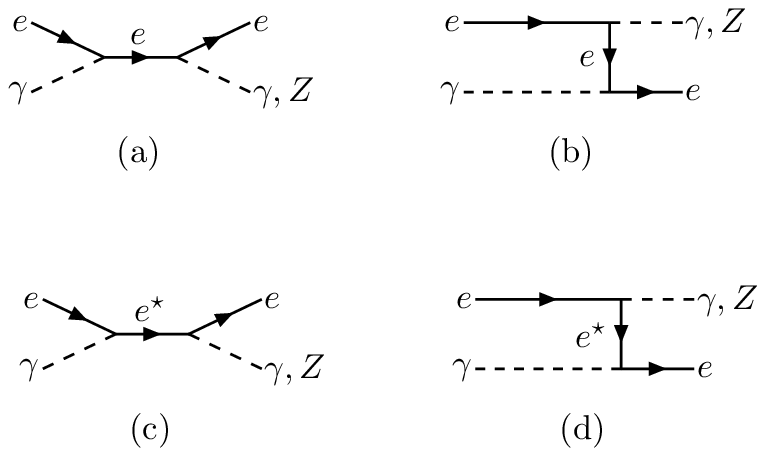,width=8cm,height=5cm}
\end{center}
\caption{Diagrams for the process $e\protect\gamma\rightarrow
e\protect\gamma,eZ$.} \label{fig2}
\end{figure}

\begin{figure}[tbp]
\begin{center}
\epsfig{file=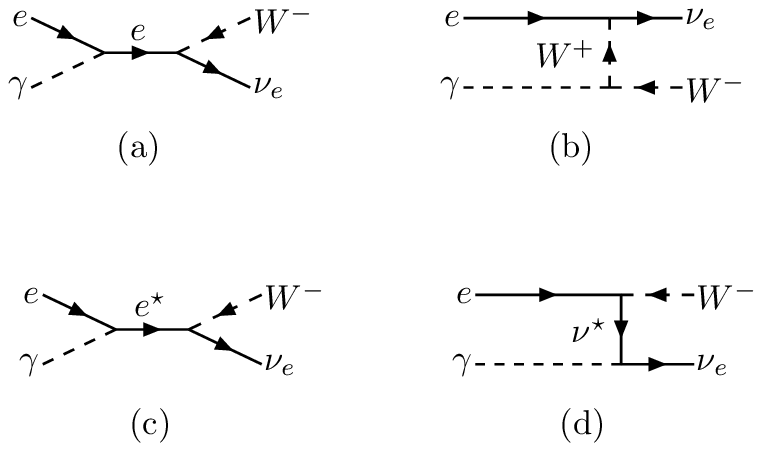,width=8cm,height=5cm}
\end{center}
\caption{Diagrams for the process $e\protect\gamma\rightarrow
\protect\nu _eW^-$.} \label{fig3}
\end{figure}

\begin{figure}[tbp]
\begin{center}
\epsfig{file=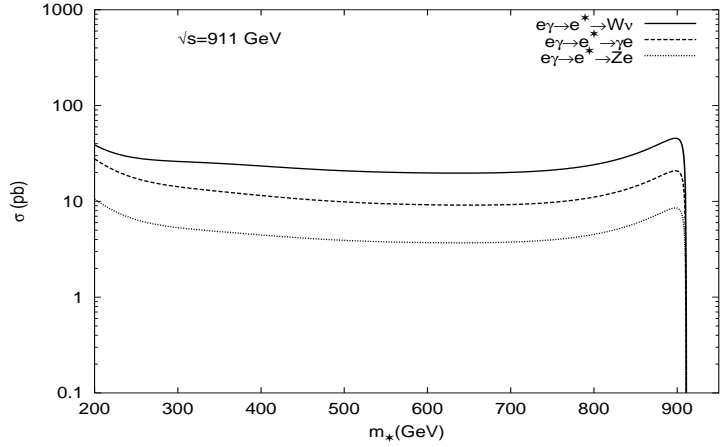,width=10cm,height=6cm}
\end{center}
\caption{The production cross sections of excited electron
depending on its mass in three different channels  at TESLA and
CLIC based $e\gamma$ colliders with $\sqrt{s}=911$ GeV. }
\label{fig4}
\end{figure}

\begin{figure}[tbp]
\begin{center}
\epsfig{file=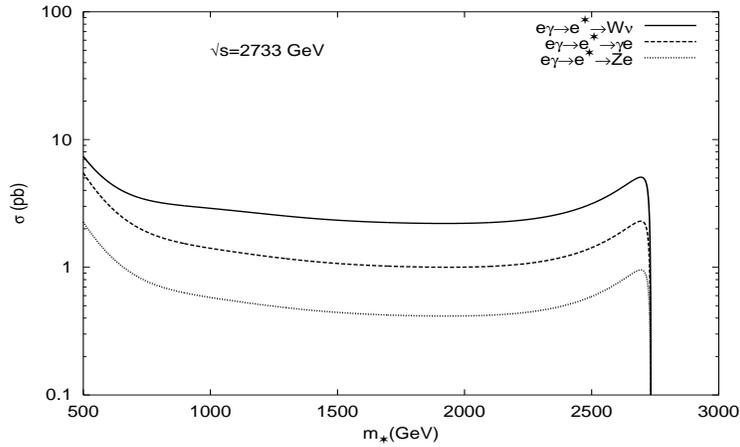,width=10cm,height=6cm}
\end{center}
\caption{The cross sections for excited electron production at
CLIC
 based $e\gamma$ colliders with $\sqrt{s}=2733$ GeV.} \label{fig5}
\end{figure}

\begin{figure}[tbp]
\begin{center}
\epsfig{file=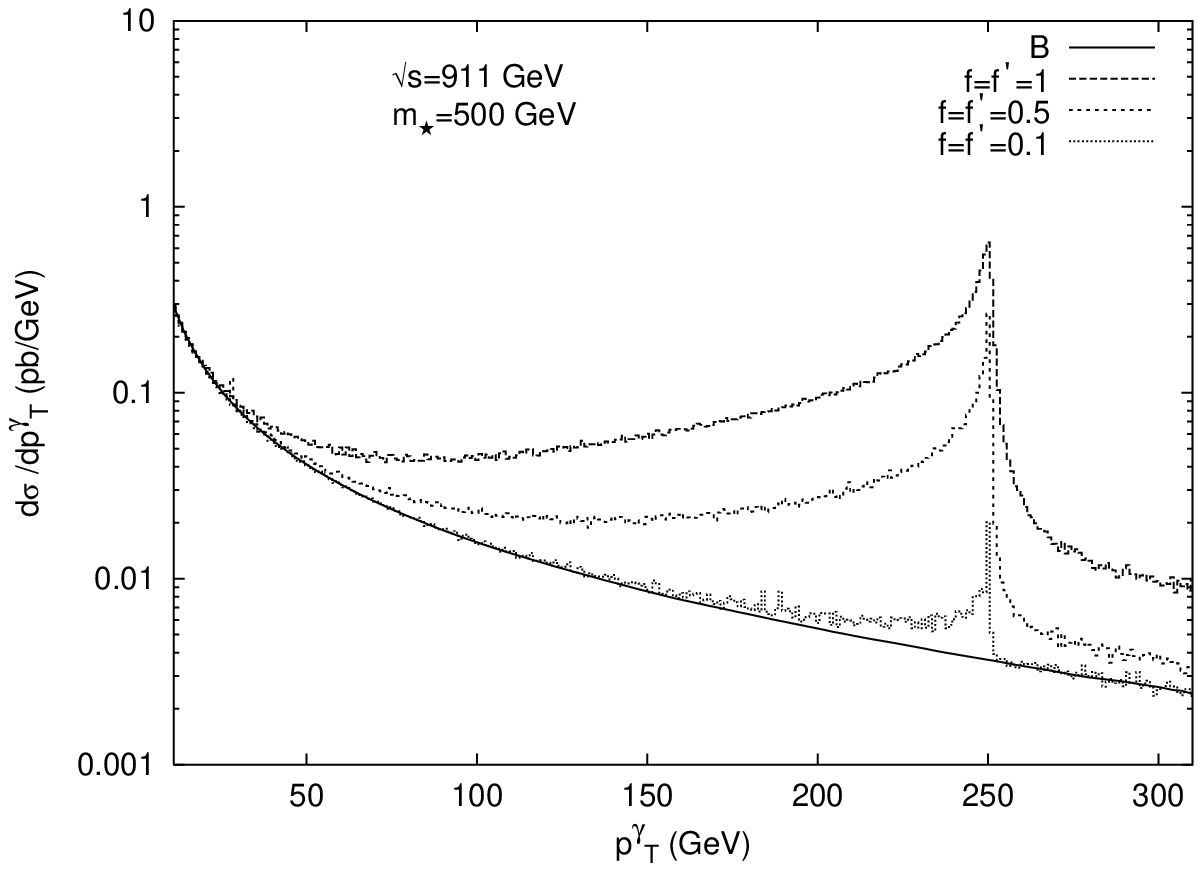,width=10cm,height=6cm}
\end{center}
\caption{Transverse momentum distribution of photon for {$e\gamma
\rightarrow e\gamma$} process according to different couplings
$f=f^\prime$ for $m_{*}=500 $ GeV at TESLA based $e\gamma$
collider with $\sqrt{s}=911$ GeV.} \label{fig6}
\end{figure}

\begin{figure}[tbp]
\begin{center}
\epsfig{file=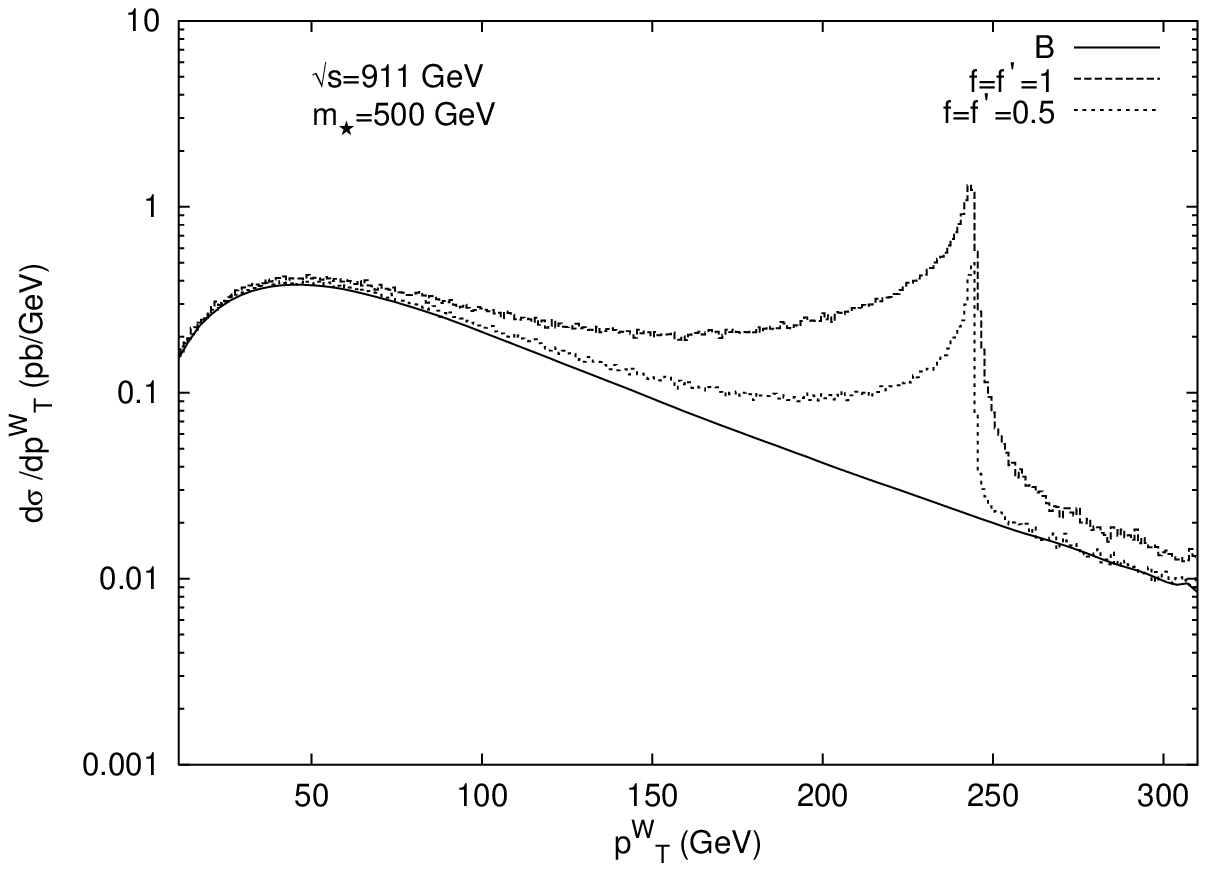,width=10cm,height=6cm}
\end{center}
\caption{Transverse momentum distribution of W boson for {$e\gamma
\rightarrow \nu W$} process according to different couplings
$f=f^\prime$ for $m_{*}=500 $ GeV at TESLA based $e\gamma$
collider with $\sqrt{s}=911$ GeV.} \label{fig7}
\end{figure}

\newpage

\begin{figure}[tbp]
\begin{center}
\epsfig{file=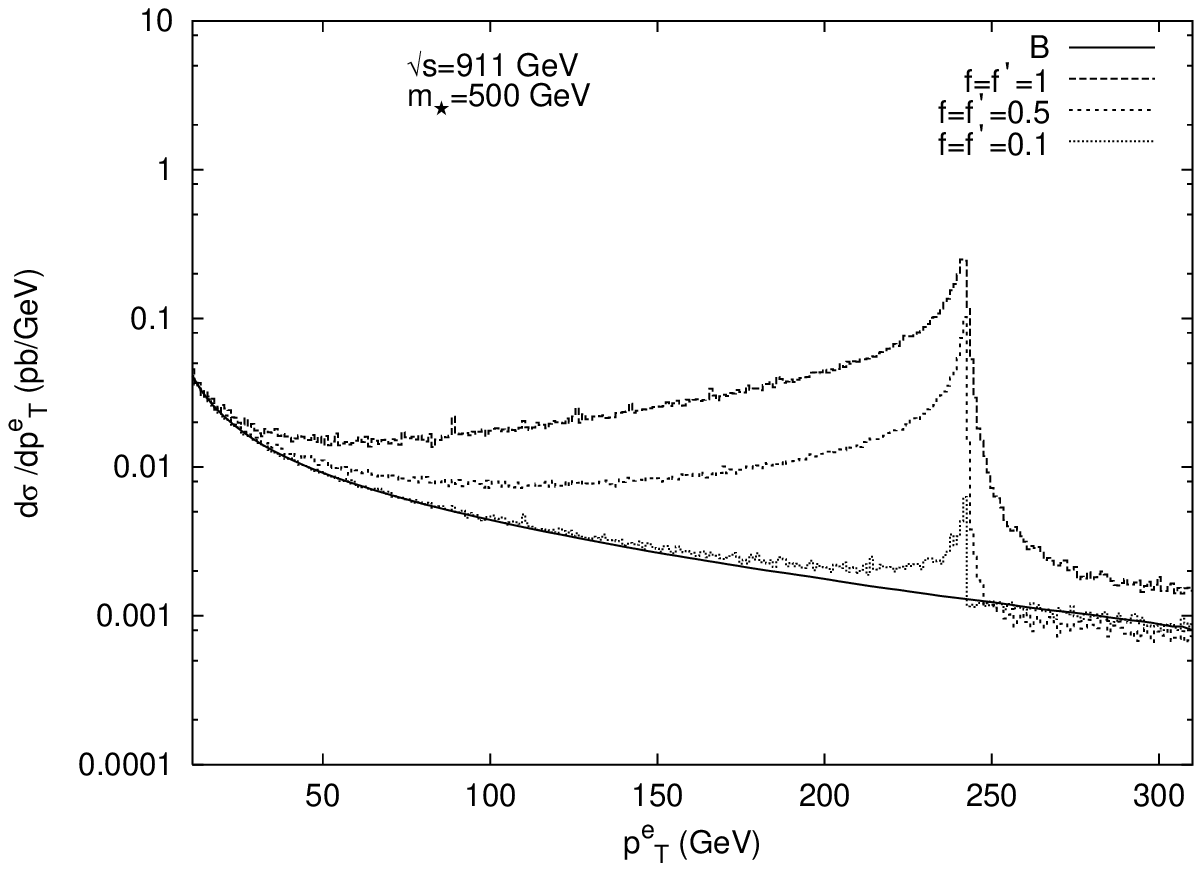,width=10cm,height=6cm}
\end{center}
\caption{Transverse momentum distribution of electron for
{$e\gamma\rightarrow eZ$} process according to different couplings
$f=f^\prime$ for $m_{*}=500 $ GeV at TESLA based $e\gamma$
collider with $\sqrt{s}=911$ GeV.} \label{fig8}
\end{figure}

\end{document}